\documentclass[usenatbib, useAMS]{mn2e}
\usepackage{graphics}
\usepackage{times}

\newcommand{\hi}{{\sc{Hi}}}                          
\newcommand{\Msun}    {\mbox{\,M\ensuremath{_\odot}}}        
\newcommand{\Mhi}     {\ensuremath{\mathrm{M_{\mbox{\scriptsize{\hi}}}}}} 
\newcommand{\sigmahi} {\ensuremath{\mathrm{\sigma_{\mbox{\scriptsize{\hi}}}}}}   
\newcommand{\Lsun}{\mbox{\,L$_\odot$}}        

\newcommand{\degrees}[1]{\ensuremath{#1^{\circ}}}

\newcommand{\Nhi}{\ensuremath{N_{\mathrm{\mbox{\scriptsize{\hi}}}}}} 

\newcommand{\kms}     {\ensuremath{\mathrm{km\,s^{-1}}}}     
\newcommand{\mJyb}{\ensuremath{\mathrm{mJy\,beam^{-1}}}}
\newcommand{\ppcsq}{\ensuremath{\mathrm{pc^{-2}}}}

\newcommand{\x}{\times}

\newcommand{\raJ}[4]{$\alpha_{\mathrm{\scriptscriptstyle J2000.0}}=#1^{\mathrm{h}}#2^{\mathrm{m}}#3^{\mathrm{s}}\!\!.#4$}

\newcommand{\raJnf}[3]{$\alpha_{\mathrm{\scriptscriptstyle J2000.0}}=#1^{\mathrm{h}}#2^{\mathrm{m}}#3^{\mathrm{s}}$}
\newcommand{\decJnf}[3]{$\delta_{\mathrm{\scriptscriptstyle J2000.0}}=#1^{\circ}#2^{\prime}#3^{\prime\prime}$}

\newcommand{\raT}[4]{$#1^{\mathrm{h}}#2^{\mathrm{m}}#3^{\mathrm{s}}\!\!.#4$}

\newcommand{\decTnf}[3]{$#1^{\circ}#2^{\prime}#3^{\prime\prime}$}

\title[Discovery of New Members in the NGC~5044 and NGC~1052 Groups]
{The Discovery of New Galaxy Members in the NGC~5044 and NGC~1052 Groups}
\author[N. P. F. M${\textstyle ^c}$Kay et. al.]
{N. P. F. M$\mathrm{^c}$Kay$^{1}$\thanks{E-mail:
npfm@astro.livjm.ac.uk},
C. G. Mundell$^{1}$\thanks{Royal Society University Research Fellow},
S. Brough$^{1}$, 
Duncan A. Forbes$^{2}$,
D. G. Barnes$^{3}$,
\newauthor
P. A. James$^{1}$,
P. Goudfrooij$^{4}$,
V. Kozhurina-Platais$^{4}$
and
R. Whitaker$^{5}$
\\ 
$^{1}$ Astrophysics Research Institute, Liverpool John Moores
University, Twelve Quays House, Egerton Wharf, \\
Birkenhead, CH41 1LD, United Kingdom \\
$^{2}$ Centre for Astrophysics and Supercomputing, Swinburne University of Technology, Hawthorn, Victoria 3122, Australia\\
$^{3}$ School of Physics, University of Melbourne, Victoria 3010,
Australia \\
$^{4}$ Space Telescope Science Institute, 3700 San Martin Drive,
Baltimore, MD 21218, USA \\
$^{5}$ Department of Physics, University of Durham, South Road, Durham, DH1 3LE, United Kingdom
}
\begin{document}

\pagerange{\pageref{firstpage}--\pageref{lastpage}} \pubyear{2002}
\maketitle
\label{firstpage}

\begin{abstract}
We present the results of neutral hydrogen (\hi{}) observations of the
NGC~5044 and NGC~1052 groups, as part of a GEMS (Group Evolution
Multiwavelength Study) investigation into the formation and evolution
of galaxies in nearby groups.  Two new group members have been
discovered during a wide-field \hi{} imaging survey conducted using
the ATNF Parkes telescope. These results, as well as those from
followup \hi{} synthesis and optical imaging, are presented here.
J$1320-1427$, a new member of the NGC~5044 Group, has an \hi{} mass of
$\Mhi=\nolinebreak[4]1.05\times10^9\Msun$ and
\Mhi/L$_B=1.65$~\Msun/\Lsun, with a radial velocity of $v=2750\kms$.
The optical galaxy is characterised by two regions of star formation,
surrounded by an extended, diffuse halo.
J$0249-0806$, the new member of the NGC~1052 Group, has 
$\Mhi=5.4\times10^8\Msun{}$, \Mhi/L$_R=1.13$~\Msun/\Lsun and
$v=1450\kms$.
The optical image reveals a low surface brightness galaxy.
We interpret both of these galaxies as irregular type, with
J$0249-0806$ possibly undergoing first infall into the NGC~1052 group.
\end{abstract}


\section{Introduction}
Galaxy groups are the most common form of galaxy association, and also
the most common environment in which galaxies are found \citep{tul87}.
The size scales of groups lie between those of galaxy pairs and rich
clusters. As such, they are an ideal environment in which to study the
relationship between the formation and evolution of galaxies, and
large scale structure.

Studies of galaxy groups and clusters at optical and infrared
wavelengths select against low surface brightness and low luminosity
objects \citep{ib97}, which are predicted to be numerous in these
environments \citep[e.g.][]{mgg+99}. These optically-faint objects are
often gas-rich and are thus more likely to be detected in neutral
hydrogen (\hi{}) surveys.
\hi{} is also a sensitive tracer of gravitational disturbances, such
as tidal interactions and galaxy mergers, with the kinematic imprint
of such encounters often being retained to distances larger than
several galactic radii \citep{mpa+95, hib00}.  Studies of the content,
distribution and kinematics of \hi{} in galaxy groups therefore
provide an important probe of ongoing galaxy interactions and mergers
and, in combination with X-ray and optical studies, provide a key
indicator of the group's evolutionary state and that of its
constituent galaxies.

The Group Evolution Multiwavelength Study (GEMS) is a multi-wavelength
investigation of the evolution of nearby galaxy groups. It compares
the X-ray, \hi{} and optical properties of a sample of groups,
selected on the availability of high-quality X-ray (ROSAT) data, with
the aim of building up a comprehensive view of their evolutionary
state and, ultimately, furthering our understanding of the processes
which drive galaxy evolution within a group environment.

As part of GEMS we have completed a sensitive, wide-area \hi{} imaging
survey of 17 nearby galaxy groups, using the Multibeam System on the
ATNF Parkes Radiotelescope\footnote{The Parkes telescope is part of
the Australia Telescope, which is funded by the Commonwealth of
Australia for operation as a National Facility managed by CSIRO}.
Full details of this survey will be given in a later paper.
With its long integration times and wide-area mapping, the GEMS \hi{}
survey complements deep targeted studies of small areas of the sky
\citep[e.g.][]{dzd+02} and less sensitive \hi{} all-sky surveys, for
example the southern HIPASS
\citep{swb+96} and northern HIJASS \citep{lbk+03} which,
although cataloguing a large number of new objects \citep{mey03}, are
less sensitive to emission from faint dwarf and LSB galaxies.

Here we report the discovery of \hi{} emission from two uncatalogued
objects in the NGC~5044 and NGC~1052 groups and present higher angular
resolution follow-up \hi{} synthesis imaging and spectroscopy, and
optical imaging of these objects. In this paper we use distances of
33~Mpc and 20~Mpc to the NGC~5044 and NGC~1052 groups, respectively,
taken from \citet{op04} in which a value of $H_0 = 70$~\kms Mpc$^{-1}$
was used.

\section{Observations and data reduction}
\subsection{Parkes observations}

Parkes observations were carried out on 2001 January
7--\nolinebreak[4]10 and July 14--18.  The area imaged for each target
group comprised a $\degrees{5.5} \times \degrees{5.5}$ field centred
on the group position and corresponding to a projected area ranging
from $0.9\times0.9$~Mpc to $3\times3$~Mpc, depending on the distance
to each of the groups in the sample.
The observations and data reduction followed the method employed by
\citet{bd01}. The spectra obtained were calibrated and corrected for
bandpass on-line and formed into an image cube, composed of two
spatial dimensions with velocity information as the third, using
software developed for the HIPASS survey \citep{bsd+01}.
These final image cubes have a spatial resolution of 15.5~arcmin with
a 4~arcmin pixel size and a spectral resolution of 1.6~\kms.
The Parkes survey parameters and the properties of the image cubes are
summarised in Table~\ref{tab:pks_obs_params}.

\begin{table}
\caption{Observing parameters for the Parkes wide-field imaging survey
and properties of the resulting image cubes.}
\begin{tabular}{lcc}
\hline
& NGC~5044 & NGC~1052 \\
\hline
Date of observation & 2001-Jan-07--10 & 2001-Jan-07--10 \\
& 2001-Jul-15--17 & 2001-Jul-14--18 \\
Centre of scanned field : & & \\
\hspace{2mm}Right Ascension (J2000) 
& \raT{13}{15}{00}{0} 
& \raT{02}{40}{00}{0}\\
\hspace{2mm}Declination (J2000) 
& \decTnf{-16}{22}{48} 
& \decTnf{-08}{08}{00}\\
Angular area imaged 
& $\degrees{5.5} \times \degrees{5.5}$
& $\degrees{5.5} \times \degrees{5.5}$ \\
Frequency of observation (MHz) & 1409.0 & 1413.5 \\
Bandwidth (MHz) & 8 & 8 \\
Spectral channels & 2048 & 2048 \\
Spectral channel width (kHz) & 3.8 & 3.8 \\
Integration time (s) & 5 & 5 \\
\hline
Data cube properties: \\
Linear dimensions (pixels) & $105\x104$ & $103\x105$ \\
Velocity channels & 1024 & 1024 \\
Velocity range (\kms)& 1565--3252 & 615--2302\\
Spatial resolution  & 15.5\arcmin & 15.5\arcmin \\
Pixel size & 4\arcmin & 4\arcmin \\
Spectral resolution (\kms) & 1.6 & 1.6 \\
RMS noise level (mJy\,beam$^{-1}$) & 53.8 & 54.0 \\
\hline
\end{tabular}
\label{tab:pks_obs_params}
\end{table}

The two cubes were inspected visually for \hi{} emission, using the
{\sc karma} data visualisation package \citep{goo95}.  Potential
detections were confirmed by extracting spectra through the cube at
each candidate position, using the {\sc mbspect} task
of the {\sc miriad} data reduction and analysis package
\citep{stw+95}.  The area over which the spectra were extracted and
the amount of Hanning smoothing used were optimised in each case, to
obtain the spectrum with maximum signal-to-noise ratio.

\begin{figure*}
  \resizebox{\textwidth}{!}{
    \rotatebox{-90}{
      \includegraphics{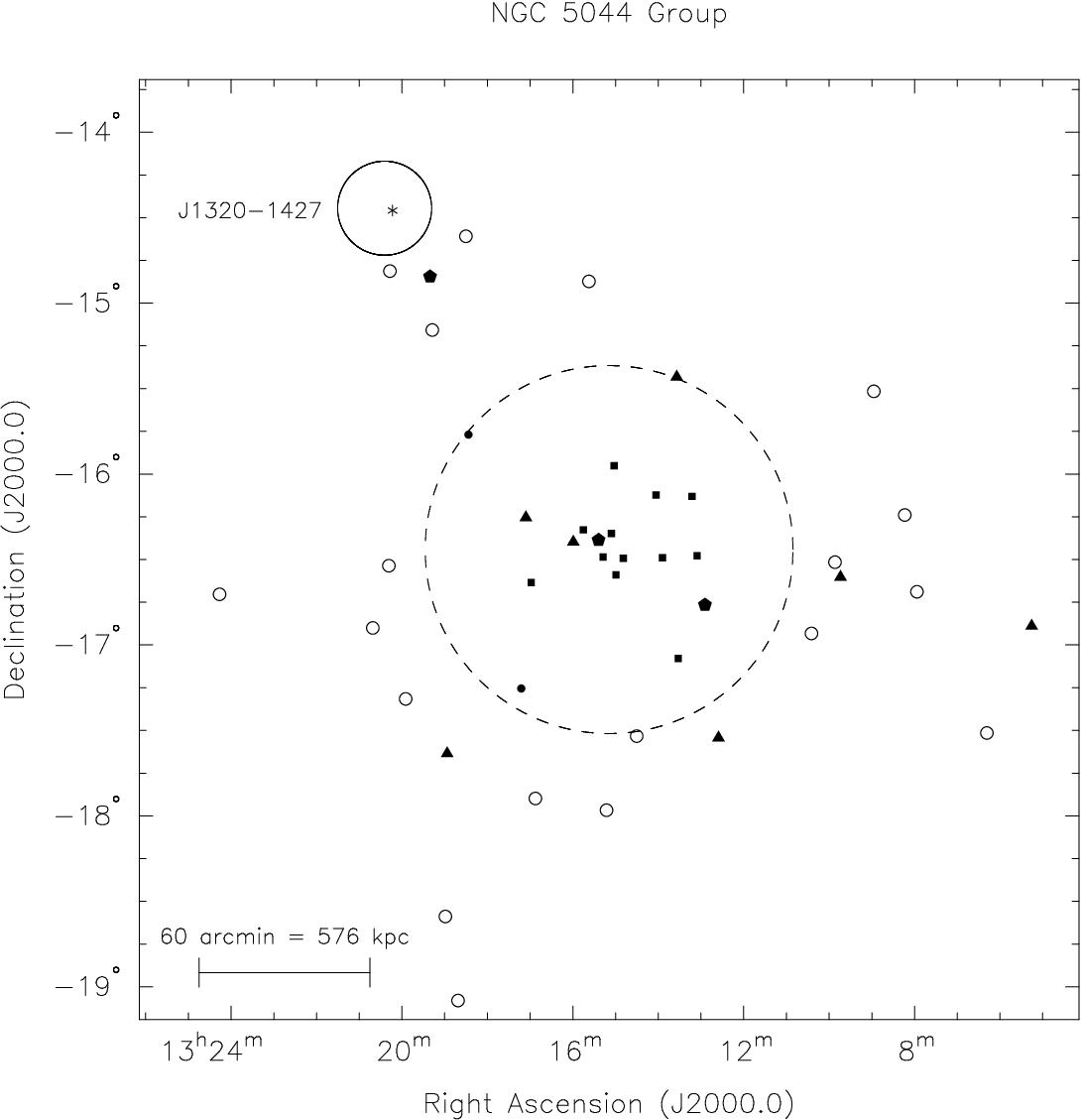}
    }
  }
\caption{Member positions in the NGC~5044 Group.
Solid symbols show the position of previously identified group
members, whilst open circles denote the positions of other probable
members (catalogued by NED) within the same projected area and with
velocities close to that of the group.
The symbol representing each member denotes the first group catalogue
to list the galaxy:
pentagon--\citet{hg82}, 
square--\citet{fs90}, 
triangle--\citet{gar93},
circle--\citet{op04}.
The position coordinates for each source are those listed by NED.
The centre of the group, at \raJ{13}{15}{09}{1}, \decJnf{-16}{26}{31},
lies very close to the galaxy NGC~5044.  The extent of the $r_{500}$
overdensity radius is denoted by a dashed circle centred on this
point.  The new group member reported in this paper, J$1320-1427$, is
in the upper left corner of the diagram, marked ``$\ast$'', and
surrounded by a circle indicating the field of view of the ATCA
observation. }
\label{fig:lgg338_pos}
\end{figure*}

In addition to emission from known group members, \hi{} emission was
detected from two positions which do not correspond to any previously
catalogued galaxies.
The location of each newly detected object, relative to its group
centre, is shown in Fig.~\ref{fig:lgg338_pos} and
Fig.~\ref{fig:lgg71_pos}.

\subsection{ATCA observations}
In order to characterise the properties and determine the more precise
locations of the newly discovered \hi{} emission regions in the
NGC~5044 and NGC~1052 groups, high-resolution $\lambda$21-cm follow-up
observations were carried out on 2002 January 8--9, using the
Australia Telescope Compact Array\footnote{The Australia Telescope is
funded by the Commonwealth of Australia for operation as a National
Facility managed by CSIRO} in its 750A configuration, which has
baselines in the range 76.5--3750.0\,m. The correlator was configured
to provide two orthogonal linear polarisations, each with a bandpass
of 8~MHz and divided into 1024~spectral channels, with the central
observing frequency tuned to the known redshifted \hi{} frequency of
the target source, yielding a velocity resolution of 1.5~km\,s$^{-1}$
per channel.  The pointing centre of each observation was chosen to be
the location of the new \hi{} emission regions detected in the Parkes
observations. These are listed in Table~\ref{tab:obs_params} along with a
summary of other observing parameters.

\begin{table*}
\caption{ATCA observing parameters and spectral line cube properties.}
\begin{tabular}{lcc}
\hline
& NGC~5044
& NGC~1052 \\
\hline
Date of observation 
& 2002-Jan-08
& 2002-Jan-09 \\
Field pointing centre:\\
\hspace{2mm}Right Ascension (J2000)
& \raT{13}{20}{24}{4}
& \raT{02}{49}{20}{8} \\
\hspace{2mm}Declination (J2000)
& \decTnf{-14}{26}{40}
& \decTnf{-08}{03}{48} \\
Primary beam (arcmin)
& 33\arcmin & 33\arcmin \\
Antenna efficiency
& 68\% & 68\% \\
Frequency of observation (MHz) & 1408 & 1414 \\
Bandwidth (MHz) & 8 & 8 \\
Spectral channels & 1024 & 1024 \\
Spectral channel width (kHz) & 7.8 & 7.8\\
Effective time on source (h) &  8.49 & 7.30  \\
Flux calibrator & PKS~B$1934-638$, & PKS~B$1934-638$ \\
& PKS~B$0823-500$\\
Phase calibrator 
& PKS~B$1308-220$
& PKS~B$0310-150$ \\
\hline
Data cube properties: \\
Linear dimensions (pixels)
&$1024 \times 1024$ 
&$1024 \times 1024$ \\
Velocity channels
& 90
& 102 \\
Velocity range (\kms)
& 2538 -- 2837
& 1174 -- 1510 \\
Channels imaged
&  451 -- 630
& 401 -- 605 \\
Spatial resolution
& 41\arcsec $\times$ 32\arcsec
& 420\arcsec $\times$ 50\arcsec \\
Pixel size
& 3\arcsec $\times$ 3\arcsec 
& 15\arcsec $\times$ 15\arcsec \\
Spectral resolution (\kms) 
& 3.3 & 3.3  \\

RMS Noise (mJy\,beam$^{-1}$) 
& 2.6    
& 3.8 \\ 
\hline
\end{tabular}
\label{tab:obs_params}
\end{table*}

The observing strategy followed the standard phase referencing
technique, consisting of alternate observations of the target source
(25~min) and a nearby phase calibrator (5~min), with this cycle
repeated over each 12~hour run.  PKS~B$1934-638$ was observed at least once
during the observation of each target, to determine the bandpass
correction and calibrate the absolute flux scale, assuming a flux
density for this source of 14.9~Jy at 1384~MHz \citep{rey94}.

The data reduction was carried out using the NRAO Astronomical Image
Processing System ({\sc aips}), with the calibration following the
standard procedure outlined in the {\sc aips} Cookbook
\citeyearpar{aip00}. The editing and calibration of the data were
carried out using a ``channel~0'' dataset, generated by averaging the
visibilities in the central 75\% of the bandpass.  

The NGC~5044 Group dataset suffered from narrow-band interference at
1.408~GHz, due to the 11th harmonic of the telescope's 128~MHz sampler
clock \citep{kil95}.  This interference was limited to a single
channel and was excised by flagging the affected channel, prior to
creation of the ``channel~0'' dataset.

Calibration of the NGC~5044 Group dataset used PKS~B$1934-638$ and the
phase calibrator PKS~B$1308-220$. Gain and phase solutions for the
calibrators were derived and applied to the spectral-line
dataset. After bandpass correction, the pure continuum dataset was
extracted (by averaging together line-free channels 100$-$450 and
630$-$950), imaged and used to further self-calibrate the target
field.  The final corrections were then applied to the spectral line
data.

Similar data editing and calibration methods were used for the
NGC~1052 Group dataset. However, due to instrumental problems, data
from antenna CA06 had to be discarded, resulting in an usable uv-range
of 0 -- 3.5~k$\lambda$ for this dataset.

On-line Doppler tracking is not used by ATCA, so calibrated spectral
line datasets were corrected for the Earth's motion using the task
{\sc cvel}. Finally, the emission in the line-free channels (100--450
and 630--950 for the NGC~5044 dataset and 200--430 and 605--850 for
the NGC~1052 dataset) was used, with the task {\sc uvlin}, to model
and subtract the continuum, in the {\em u-v} plane.

An image cube of each target field, employing 2-channel spectral
averaging (resulting in a velocity resolution of 3.3~\kms) was then
generated using {\sc imagr}, to Fourier transform and deconvolve the
calibrated, spectral-line datasets.  Natural weighting (Briggs'
robustness parameter 5) was used, to maximise sensitivity to low
surface-brightness, extended structure \citep{bri95}, and line-free
channels (i.e. those chosen to produce the continuum dataset) were
excluded from the imaging process.  Finally, a correction was applied
to the image cubes, using {\sc pbcor}, to account for primary beam
attenuation.  Spectral-line cube properties are given in
Table~\ref{tab:obs_params}.

The minimum detectable \hi{} mass per channel in the NGC~5044 cube
($3\sigma$ detection), of an object of angular size equal to that of
the synthesised beam and located at the pointing centre, is
$9\x10^6$~\Msun.  Similarly, a $3\sigma$~detection of an object with
the largest angular extent imageable by the array ($\sim4$~arcmin) is
$4\x10^8$~\Msun.

For the NGC~1052 dataset, the minimum detectable \hi{} mass per
channel is $5\x10^6$\Msun, assuming a $3\sigma$~detection at the
pointing centre, of an object the size of the synthesised
beam. Similarly, a $3\sigma$~detection of an object with the largest
angular extent imageable by the array ($\sim4$~arcmin) is
$1\x\nolinebreak[4]10^7$~\Msun.

\subsection{Optical observations}
Examination of images from the Second Digitized Sky Survey
\citep[DSS\,II --][]{lm94} revealed faint optical emission coincident
with the \hi{}-detected objects in the NGC~5044 and NGC~1052 groups
(see Section~\ref{sec:results}). Subsequently, follow-up broad-band
images were obtained using the Wide Field Imager (WFI) on the MPG/ESO
2.2-m telescope in 2001 August 7-10.  Conditions were photometric with
$\sim$1~arcsec seeing. The exposure times were $5 \x 120$~s
and $4 \x 300$~s respectively, for the {\em B}- and 
{\em R}-band observations of the NGC~5044 group, and 
$5 \x 120$~s for the {\em R}-band observation of the NGC~1052 
group.

Data were taken at several dither positions, in order to
eliminate the gaps between the individual CCD chips of the WFI
instrument, when mosaicing the dithered images together after basic
reduction of the individual images. The data reduction was carried out
using IRAF software as well as dedicated IDL scripts, and galaxy
photometry was performed using {\sc Sextractor} (\citealp{ba96}; see
\citealp{mrf+03} for further details).  The {\sc Sextractor}
detection threshold for galaxies detected in the mosaiced {\em R}-band
image of the NGC5044 field was $R = 22.85$.

\section{Results}
\label{sec:results}
Regions of \hi{} emission from positions previously uncatalogued in
both optical and radio positional databases
(e.g. the NASA-IPAC Extragalactic Database (NED\footnote{The
NASA/IPAC Extragalactic Database (NED) is operated by the Jet
Propulsion Laboratory, California Institute of Technology, under
contract with the National Aeronautics and Space Administration.}))
are detected in the NGC~5044 and NGC~1052 groups.  In the NGC~5044
Group, \hi{} is detected $\sim$\degrees{2.5} North-East of the group
centre, whilst in the NGC~1052 Group \hi{} emission is detected 14
arcmin South of the edge-on galaxy NGC~1110, itself located
\degrees{2} from the group centre. Optical emission is also detected
from these two \hi{}-emission regions.

\begin{table}
\caption{Group parameters for the NGC~5044 and NGC~1052
groups, from \citet{op04}.}
\begin{tabular}{lcc}
\hline
& NGC~5044
& NGC~1052 \\
\hline
Group centre:\\
\hspace{2mm}Right Ascension (J2000) 
& \raT{13}{15}{09}{1}
& \raT{02}{40}{35}{3} \\
\hspace{2mm}Declination (J2000)   
& \decTnf{-16}{26}{31}
& \decTnf{-08}{13}{08} \\
Systemic velocity (\kms{}) 
& $2518 \pm 100$  
& $1366 \pm 41$ \\
Velocity dispersion (\kms{}) 
& $426 \pm 74$  
& $91 \pm 35$ \\
Distance to group (Mpc) 
& 33     
& 20 \\  
$r_{500}$ (kpc)
& 620
& 360 \\
\hline
\end{tabular}
\label{tab:group_params}
\end{table}

\subsection{NGC~5044 Group: J1320--1427}
\label{sec:pks_5044}
\subsubsection*{\hi{} emission}

\begin{figure*}
\resizebox{\textwidth}{!}{
  \includegraphics*[40,24][550,528]{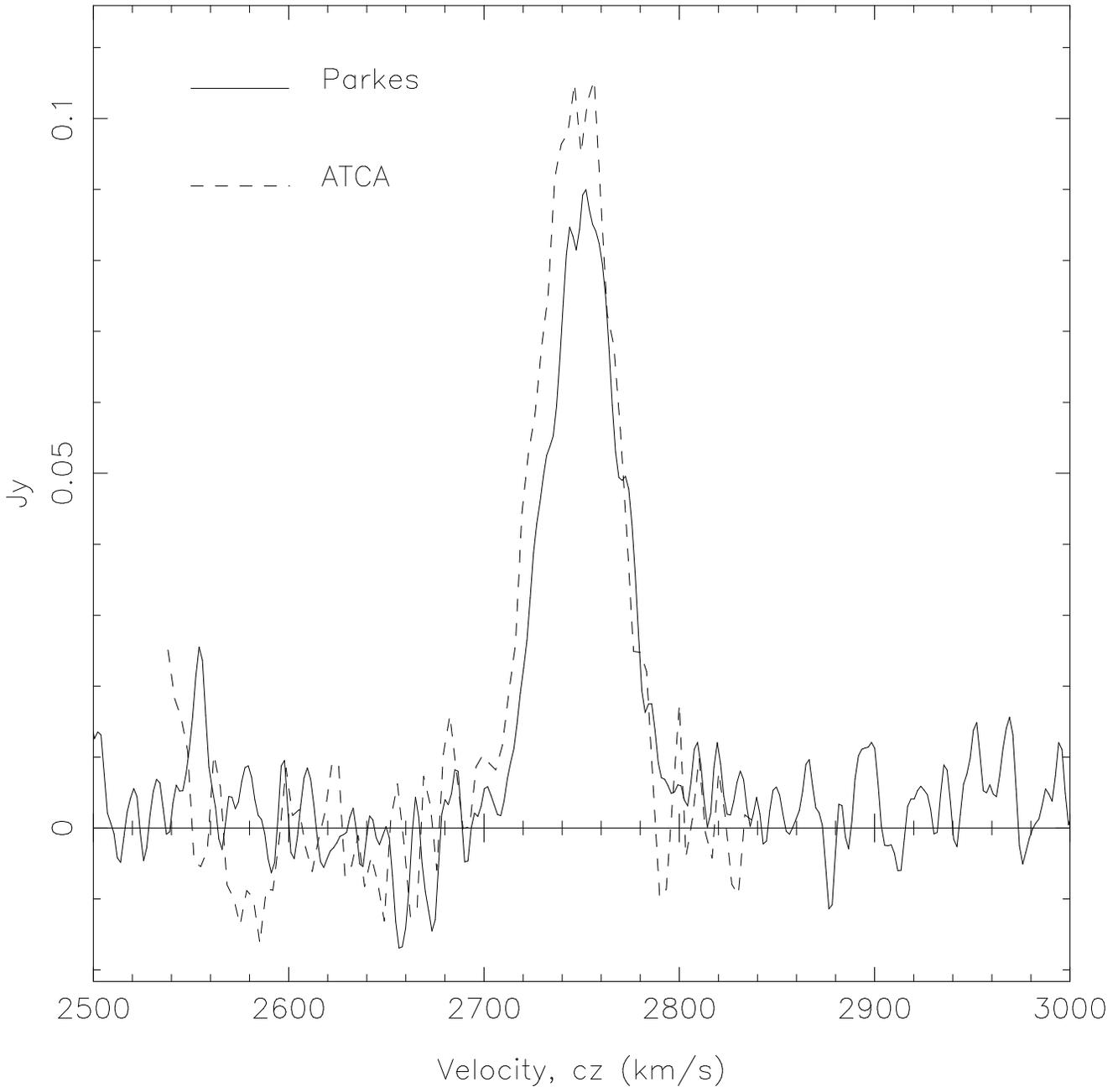}
}
\caption{NGC~5044 Group: Parkes (solid) and ATCA (dashed) \hi{} spectra of 
J$1320-1427$.
The Parkes spectrum was produced by averaging over a spatial box of
width 3 pixels (12~arcmin) and Hanning smoothing over 7 velocity
channels. A second-order polynomial has been used to correct the
non-zero baseline.
The ATCA spectrum was taken over a spatial region whose boundaries are
defined by the $3\sigma$ detection level (a column density of 
$2.0\times10^{20}$~atoms~cm$^{-2}$).
}
\label{fig:n5044spectra}
\end{figure*}

Fig.~\ref{fig:n5044spectra} shows the Parkes spectrum of the
previously uncatalogued source of \hi{} emission, discovered
$\sim$\degrees{2.5} North-East of the NGC~5044 Group centre (see
Table~\ref{tab:group_params}), to which we refer as ``J$1320-1427$''
after its refined position. The spectrum was extracted using a spatial
box width of 5~pixels and Hanning smoothing over 7 channels; a
second-order polynomial was fitted to line-free channels and
subtracted to correct the non-zero baseline.  The line has a width, at
50\% intensity, of $W_{50}=47$~\kms{}.  The integrated
\hi{} flux is $\int\mathrm{S}~d\mathrm{V}=4.1\pm0.1$~Jy\,km\,s$^{-1}$
which, assuming optically thin emission and a distance to J$1320-1427$
of 33~Mpc (the adopted distance to the NGC~5044 Group), implies a
total
\hi{} mass of $(1.05\pm0.03)\times10^9\Msun$.

\begin{figure*}
  \resizebox{\textwidth}{!}{
    \includegraphics*[13,39][547,551]{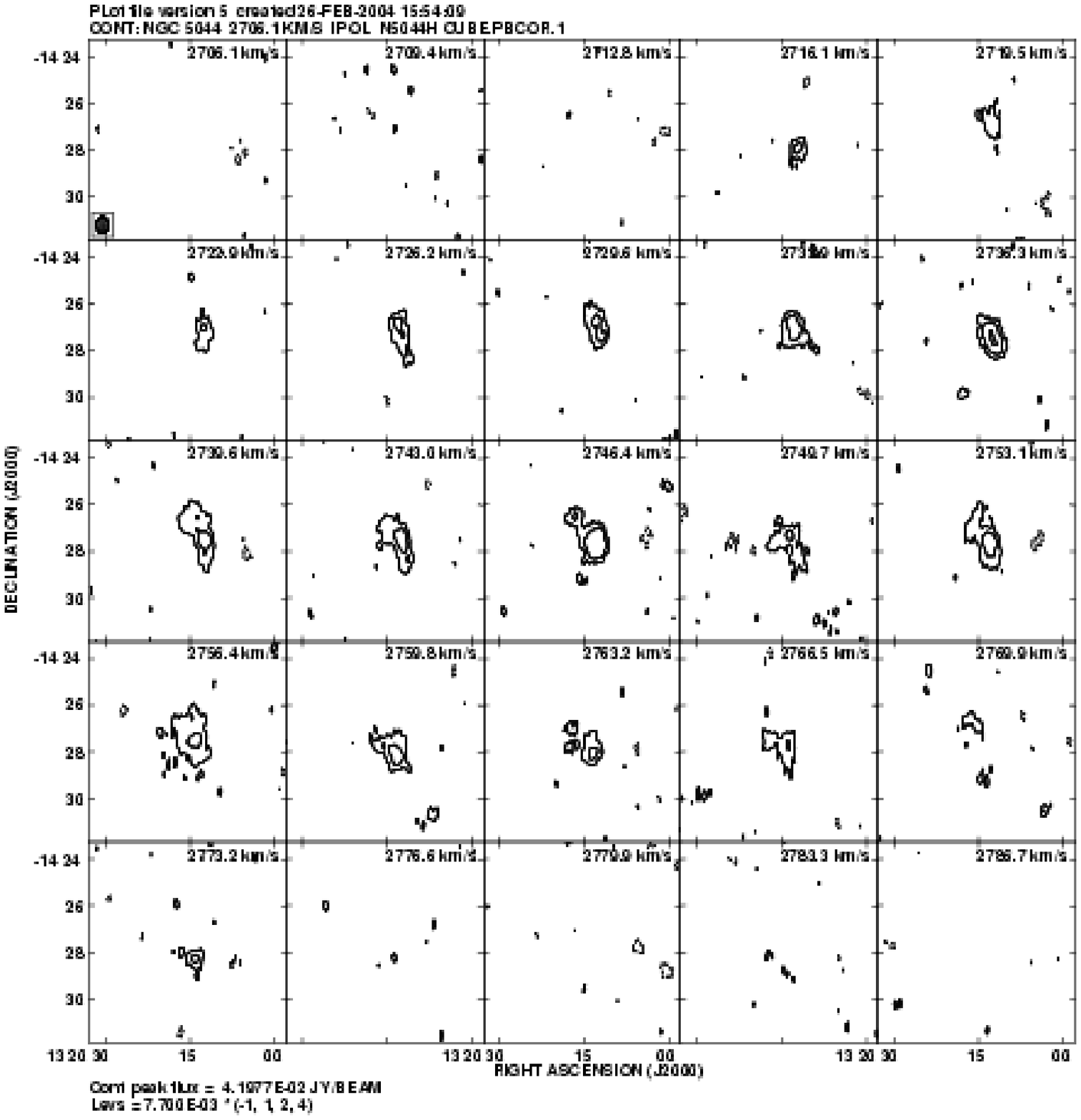}
  }
\caption{NGC~5044 Group: Channel maps showing \hi{} emission from the new group
member J$1320-1427$. The peak of the emission is 33~\mJyb{}, at a
velocity of 2736.3~\kms{}.  The contour levels represent flux levels of
$-3\sigma$ (dashed), $3\sigma$, $6\sigma$ and $12\sigma$, where the
noise level $\sigma = 2.6$~mJy~beam$^{-1}$. 
A $3\sigma$ detection corresponds to an \hi{} column density of
$\Nhi=3.3\times10^{19}$~cm$^{-2}$ (or 0.26~\Msun\ppcsq) per channel.
The central velocity
(heliocentric, optical definition) of each channel is shown in the top
right of each map, and the size of the synthesised beam ($41\arcsec
\times 32\arcsec$) is shown in the bottom-left of the map of the
lowest velocity channel.  }
\label{fig:n5044hi_chan_maps}
\end{figure*}

ATCA imaging of J$1320-1427$ reveals \hi{} emission from a region
centred on \raJnf{13}{20}{13}, \decJnf{-14}{27}{32}, with a
deconvolved size of $65\arcsec\times125\arcsec$ and a velocity range
of 2716--2773~\kms{}.
Channel maps for this object are shown in Fig.~\ref{fig:n5044hi_chan_maps}.

\begin{figure*}
  \resizebox{\textwidth}{!}{
    \includegraphics*[16,47][551, 648]{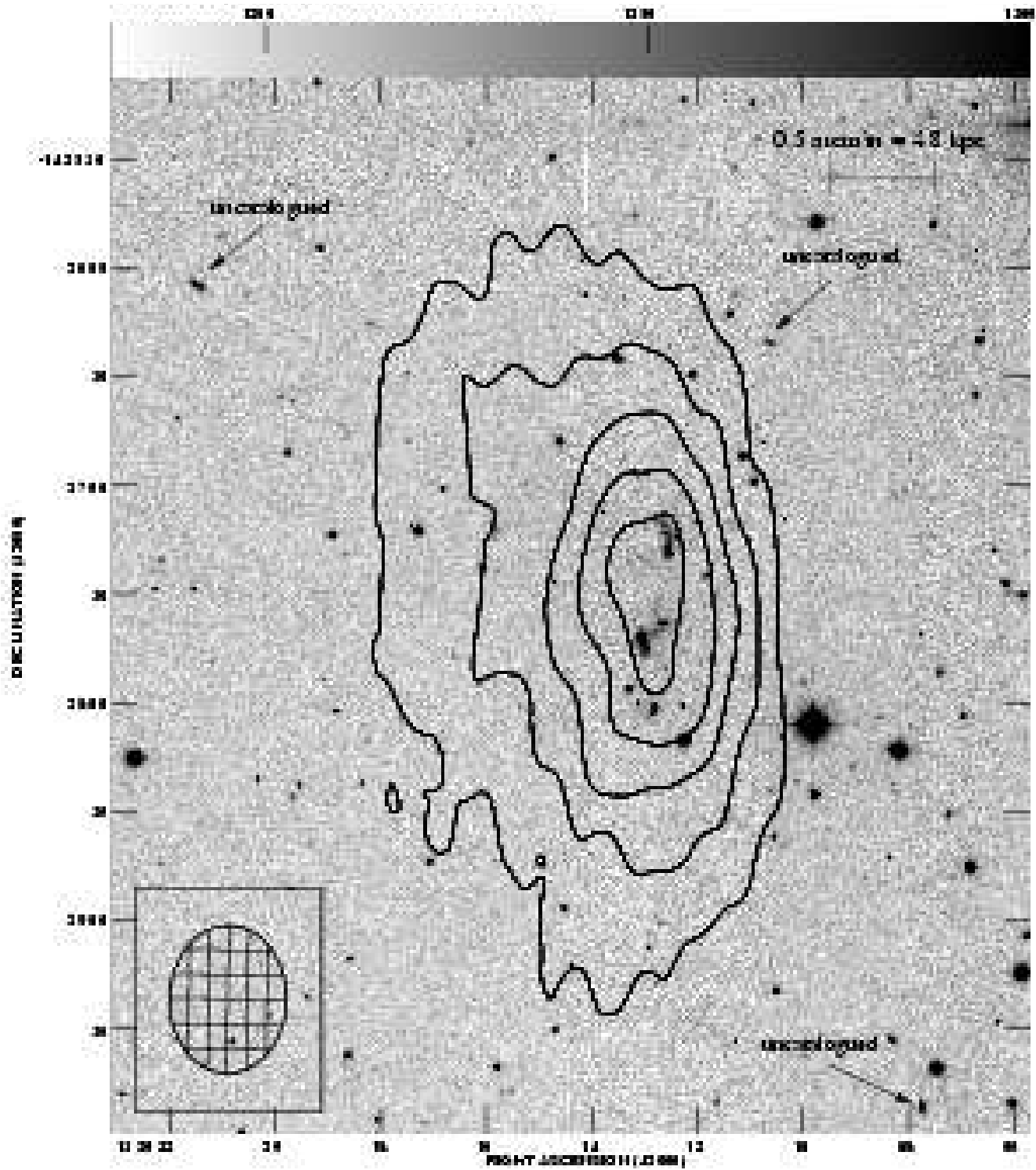}
}
\caption{J$1320-1427$ in the NGC~5044 Group: Total \hi{} intensity contours
superimposed on to the MPG/ESO 2.2~m WFI {\em R}-band optical image. 
The peak column density of the source is $1.1 \times 10^{21}$~atoms\,cm$^{-2}$ 
(9.0~\Msun\ppcsq{}) and contour levels are set to $3\sigma$, $6\sigma$,
$9\sigma$, $12\sigma$, $15\sigma$.
These represent column densities of 
1, 2, 3, 4, and 5 times $2.0 \times 10^{20}$~atoms\,cm$^{-2}$ 
(equivalent to 1, 2, 3, 4, and 5 times 1.6~\Msun\ppcsq{}).
The \hi{} map has been rescaled to the same pixel size and coordinate
grid as the optical data.  The ellipse in the lower-left of the image
indicates the size of the synthesised beam ($41\arcsec \times
32\arcsec$) and the scale bar at the top-right shows the linear scale
of the image (1~arcsec~=~160~pc at a distance of 33~Mpc).  At least
3 other galaxies can be seen in the optical image, but none of these
is identified in the standard catalogues.  }
\label{fig:n5044_m0}
\end{figure*}

\begin{figure*}
  \resizebox{\textwidth}{!}{
    \includegraphics*[18,64][551,641]{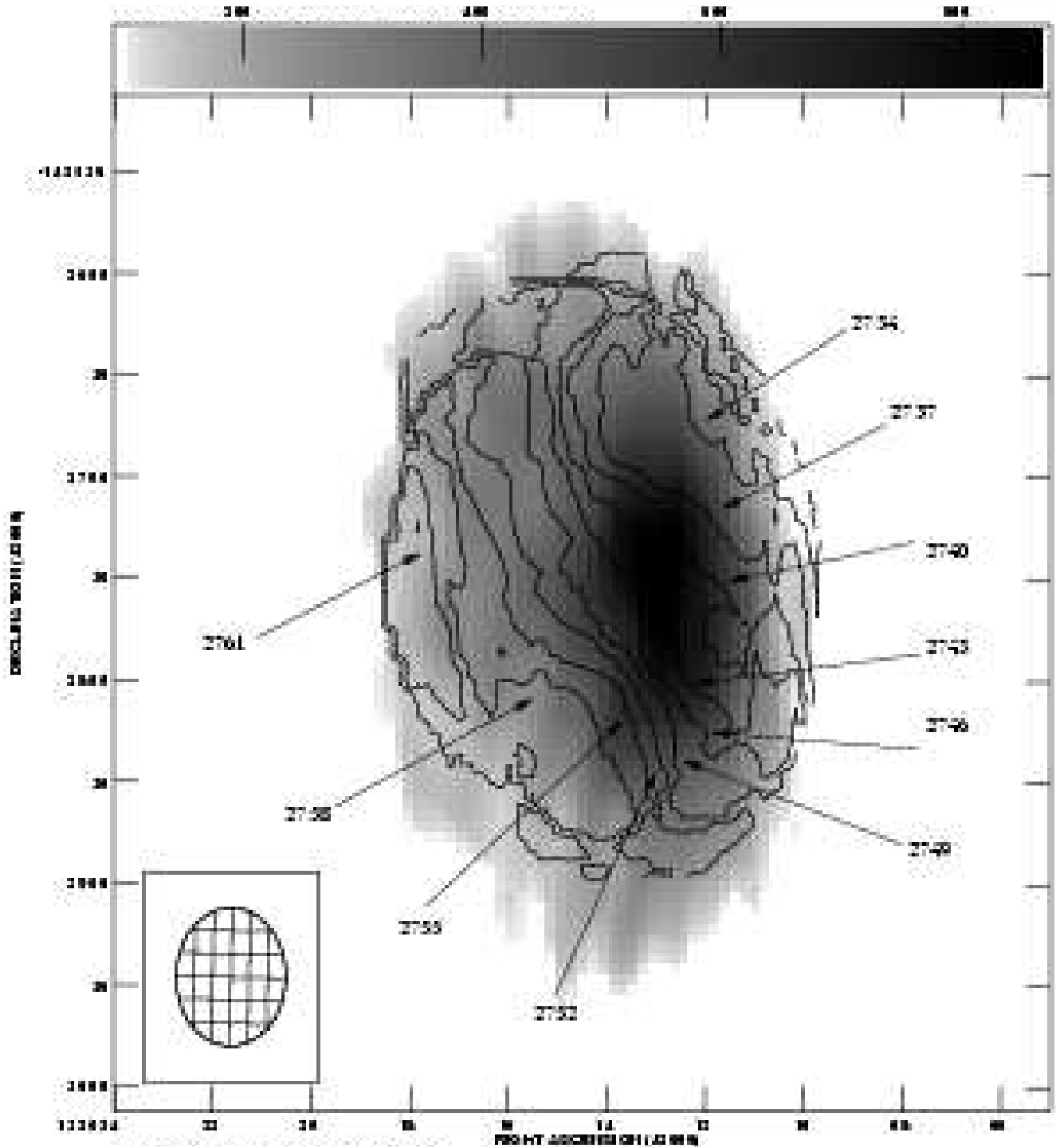}
  }
\caption{J$1320-1427$ in NGC~5044 Group: \hi{} isovelocity contours superimposed
on to the greyscale \hi{} intensity map. As indicated in the figure,
the velocity contours cover the range 2731 -- 2761~\kms{} in regular
3~\kms{} increments. The ellipse in the lower-left of the image
indicates the size of the synthesised beam ($41\arcsec \times
32\arcsec$).
\newline
\newline
}
\label{fig:n5044_m0m1}
\end{figure*}

Maps of line intensity integrated over velocity
($0^{\mathrm{th}}$~moment, Fig.~\ref{fig:n5044_m0}) and
intensity-weighted velocity ($1^{\mathrm{st}}$~moment,
Fig.~\ref{fig:n5044_m0m1}) were produced using the Brinks {\em
``conditional blank''} method \citep{tbs93} where a mask, created by
excluding fluxes below $2\sigma$ on each spatially smoothed channel
map, was used to filter noise from the full resolution image cube,
prior to calculating the moments with {\sc xmom}.
The total \hi{} intensity map (Fig.~\ref{fig:n5044_m0}) shows the
emission to be elongated in the north-south direction, with emission
extending to the north-east. The $3\sigma$ contour of this map was
used to define the spatial area over which the average spectrum, shown
in Fig.~\ref{fig:n5044spectra}, was taken.

We measure a peak \hi{} column density of
$1.1~\times~10^{21}$~atoms\,cm$^{-2}$ (9.0~\Msun\ppcsq{}), 
which is a lower limit of this quantity, due to the large beam size.
The integrated total \hi{} intensity is 
$\int\mathrm{S}~d\mathrm{V}=4.3\pm0.2$~Jy\,km\,s$^{-1}$, from which we
calculate an \hi{} mass $\Mhi=(1.10\pm0.05)\times10^9$~\Msun. The
projected size of the \hi{} emission, measured to a column density of
1.6~\Msun\ppcsq{} ($3\sigma$), is $11 \times 20$~kpc.

Despite the narrow range in velocity covered by the \hi{} emission
($W_{50}=47$~\kms{} - see Fig.~\ref{fig:n5044spectra}), the velocity
field (Fig.~\ref{fig:n5044_m0m1}) is relatively well-ordered, showing
a smooth change in velocity of $\sim$30~\kms{} over $\sim$1.5 arcmin
(14~kpc) from the north-west to south-east.

Several background radio continuum sources lie within the field of
view but no continuum emission above the $3\sigma$ level of
0.3~\mJyb{} was detected at the position of J$1320-1427$.

\subsubsection*{Optical data}
\label{sec:opt_lgg338}
\begin{table*}
\caption{J$1320-1427$ in NGC~5044 Group: Measured and derived optical 
properties. The columns are as follows: {\sc (i -- iv)} - each of the
bright components labelled 1 -- 4 in Fig.~\ref{fig:n5044_opt}; {\sc
(v)} - the low surface-brightness (LSB) component only; {\sc (vi)} -
the combined properties of all five components (i.e. 1, 2, 3, 4 and
LSB).
Absolute magnitudes, M$_B$ and M$_R$, and luminosities, L$_B$ and
L$_R$, are corrected for extinction. Estimates of Galactic extinction,
A$_B$ and A$_R$, have been derived as per
\citet{sfd98}. A$_B$(intrinsic) and A$_R$(intrinsic) are the
extinction, for the B and R bands respectively, derived from the total
\hi{} content of J$1320-1427$.
The surface brightness calculation uses the ellipse shown in
Fig.~\ref{fig:n5044_opt} as the definition of the extent of the
LSB component. 
}
\begin{tabular}{l*{6}{c}}
\hline
\multicolumn{7}{c}{J$1320-1427$} \\
                                & Component 1      & Component 2      & Component 3      & Component 4     & LSB            & Total           \\ 
                                & \sc(i)           & \sc(ii)          & \sc(iii)         & \sc(iv)         & \sc(v)         & \sc(vi)         \\ 
\hline                                                                                                                                                                            
m$_B$                     (mag) & $20.15$          & $18.42$          &  $24.39$         & $18.66$         & $18.27$        & $17.17$         \\ 
B $-$ R                   (mag) & $1.42$           &  $0.32$          &  $3.90$          & $0.16$          & $1.89$         & $1.22$          \\ 
                                                                                                                                                                                  
$\mathrm{A_B}$(Galactic)  (mag) & 0.37             & 0.37             & 0.37             & 0.37            & 0.37           & -               \\ 
$\mathrm{A_R}$(Galactic)  (mag) & 0.23             & 0.23             & 0.23             & 0.23            & 0.23           & -               \\ 
$\mathrm{A_B}$(intrinsic) (mag) & 0.68             & 0.85             & 0.57             & 0.82            & 0.63           & -               \\ 
$\mathrm{A_R}$(intrinsic) (mag) & 0.38             & 0.48             & 0.32             & 0.46            & 0.35           & -               \\ 
                                                                                                                                                                                   
M$_B$                     (mag) & $-13.49$         & $-15.39$         & $-9.14$          & $-15.12$        & $-15.32$       & $-16.54$        \\ 
M$_R$                     (mag) & $-14.47$         & $-15.20$         & $-12.65$         & $-14.78$        & $-16.79$       & $-17.25$        \\ 
                                                                                                                                                                                  
$\mu_B$   (mag arcsec$^{-2}$)   & -                & -                & -                & -               & 26.91          & 25.81           \\ 
$\mu_R$   (mag arcsec$^{-2}$)   & -                & -                & -                & -               & 25.02          & 25.59           \\ 
                                                                                                                                                                                  
L$_B$                 (\Lsun)   & $3.84 \x 10^7$   & $2.21 \x 10^8$   & $6.98 \x 10^5$   & $1.72 \x 10^8$  & $2.07 \x 10^8$ & $6.39 \x 10^8$  \\ 
L$_R$                 (\Lsun)   & $3.16 \x 10^7$   & $6.19 \x 10^7$   & $5.92 \x 10^6$   & $4.21 \x 10^7$  & $2.68 \x 10^8$ & $4.10 \x 10^8$  \\ 
                                                                                                                                                                                  
\Mhi / L$_B$      (\Msun/\Lsun) & -                & -                & -                & -               & 1.37           & 1.64            \\ 
\Mhi / L$_R$      (\Msun/\Lsun) & -                & -                & -                & -               & 1.05           & 2.56            \\ 
\hline
\end{tabular}
\label{tab:n5044_opt}
\end{table*}

\begin{figure*}
  \begin{minipage}[c]{0.49\linewidth}
    \begin{center}
       \resizebox{!}{8cm}{
         \includegraphics*[29,29][540,540]{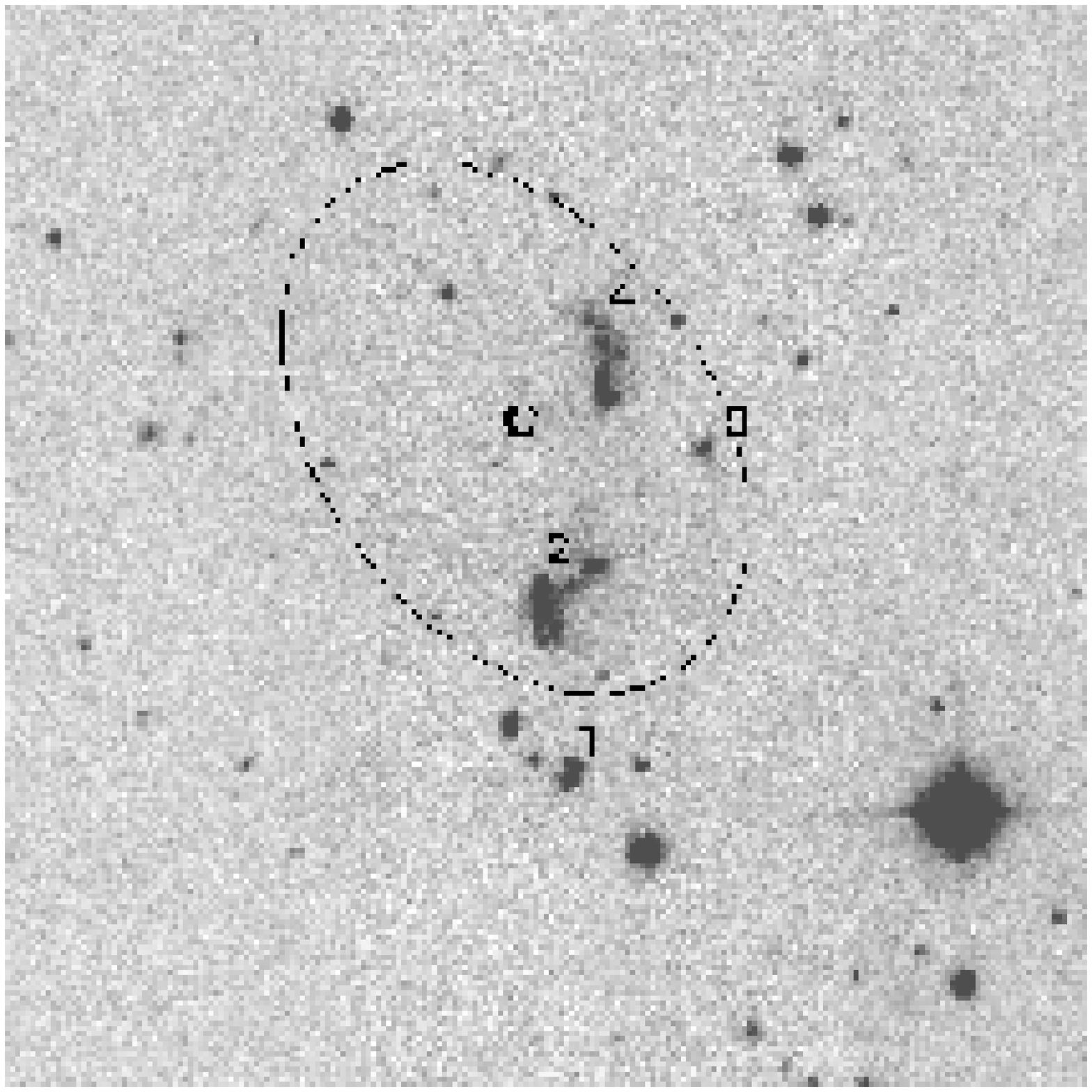}
       }
    \end{center}
  \end{minipage} \hfill
  \begin{minipage}[c]{0.49\linewidth}
    \begin{center}
       \resizebox{!}{8cm}{
         \includegraphics*[22,23][428,427]{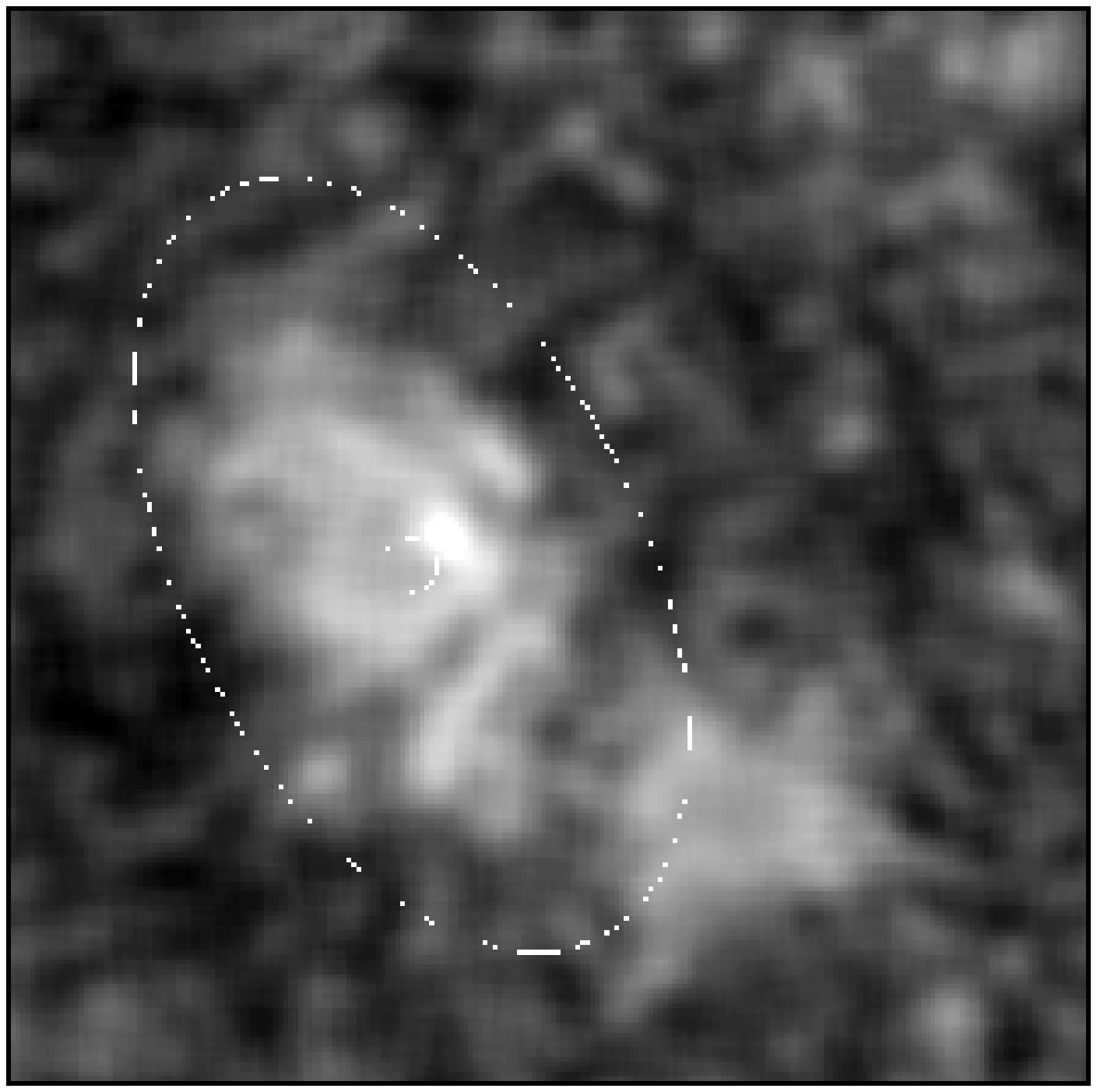}
       }
    \end{center}
  \end{minipage}
\caption{J$1320-1427$ in NGC~5044 Group: Left --- MPG/ESO 2.2~m WFI 
{\em R}-band image. The ellipse illustrates the extent of the low
surface brightness component, with its centre marked C. The brightest
components are numbered 1 -- 4.
Right --- Convolved image showing the low surface brightness (LSB)
component, with the {\sc Sextractor} ellipse overlaid. The apparent
emission to the south-west of the LSB component is a residue of
imperfect masking of features caused by the nearby saturated star,
clearly visible in the left panel.
}
\label{fig:n5044_opt}
\end{figure*}

Our follow-up {\em B} and {\em R}-band images (see
Fig.~\ref{fig:n5044_opt}) reveal knots of optical emission; two
spatially-extended sources straddle the position of \hi{} maximum,
while two further bright regions lie, one south and one west of the
main peaks.
The two main optical components (labelled 2 and 4 in
Fig.~\ref{fig:n5044_opt}) are very blue, with ({\em B}$-${\em R})
values of 0.32 and 0.16 magnitudes respectively, suggesting that star
formation is occurring within these regions.

Applying a filter to remove the high surface-brightness objects in the
{\em R}-band image and convolving the result with an exponential
kernel, revealed a very low-surface brightness halo extending
north-east of the two main optical components, towards the extended
\hi{} emission.  Fig.~\ref{fig:n5044_opt} shows the {\em R}-band image
with the four main optical components identified, and the convolved
image showing the LSB halo. The extent of the faint halo is
illustrated with an ellipse.  The mean sky background level in the
R-band mosaic image was 20.24~mag\,arcsec$^{-2}$, while the peak
surface brightness of the LSB component was 23.77~mag\,arcsec$^{-2}$.
In general, the convolution of the image with an appropriate
exponential kernel allowed us to detect galaxies with peak surface
brightnesses of about 4.25~magnitudes below the background.  The
detection of the LSB component is therefore significant, being
0.75~mag brighter than the detection limit.

The measured and derived optical properties of the system are shown in
Table~\ref{tab:n5044_opt}.  Columns numbered {\sc(i -- iv)} refer to
the four bright optical components, and column {\sc(v)} to the low
surface brightness halo (LSB). Column {\sc(vi)} shows the combined optical
properties of all objects within the extent of the halo
(components 2, 4 and LSB).  The Galactic extinction corrections in
{\em B} and {\em R}, $A_B$ and $A_R$, were obtained
as per \citet{sfd98}.

The upper limit for the extinction intrinsic to J$1320-1427$ was
calculated from the average \hi{} column density over each component,
using the relations $A_B=8.62\x10^{-22}\Nhi$ \citep{sd87} and
$A_R=A_B/1.770$ \citep{bm98}.  The surface brightnesses ($\mu_B$ and
$\mu_R$) of the LSB component only, and the total surface brightness
of the optical object were calculated assuming that the optical extent
is defined by the ellipse shown in Fig.~\ref{fig:n5044_opt}.

\subsection{NGC~1052 Group: J0249--0806}
\label{sec:pks_1052}
\subsubsection*{\hi{} emission}
\begin{figure*}
\resizebox{\textwidth}{!}{
  \includegraphics*[40,24][550,527]{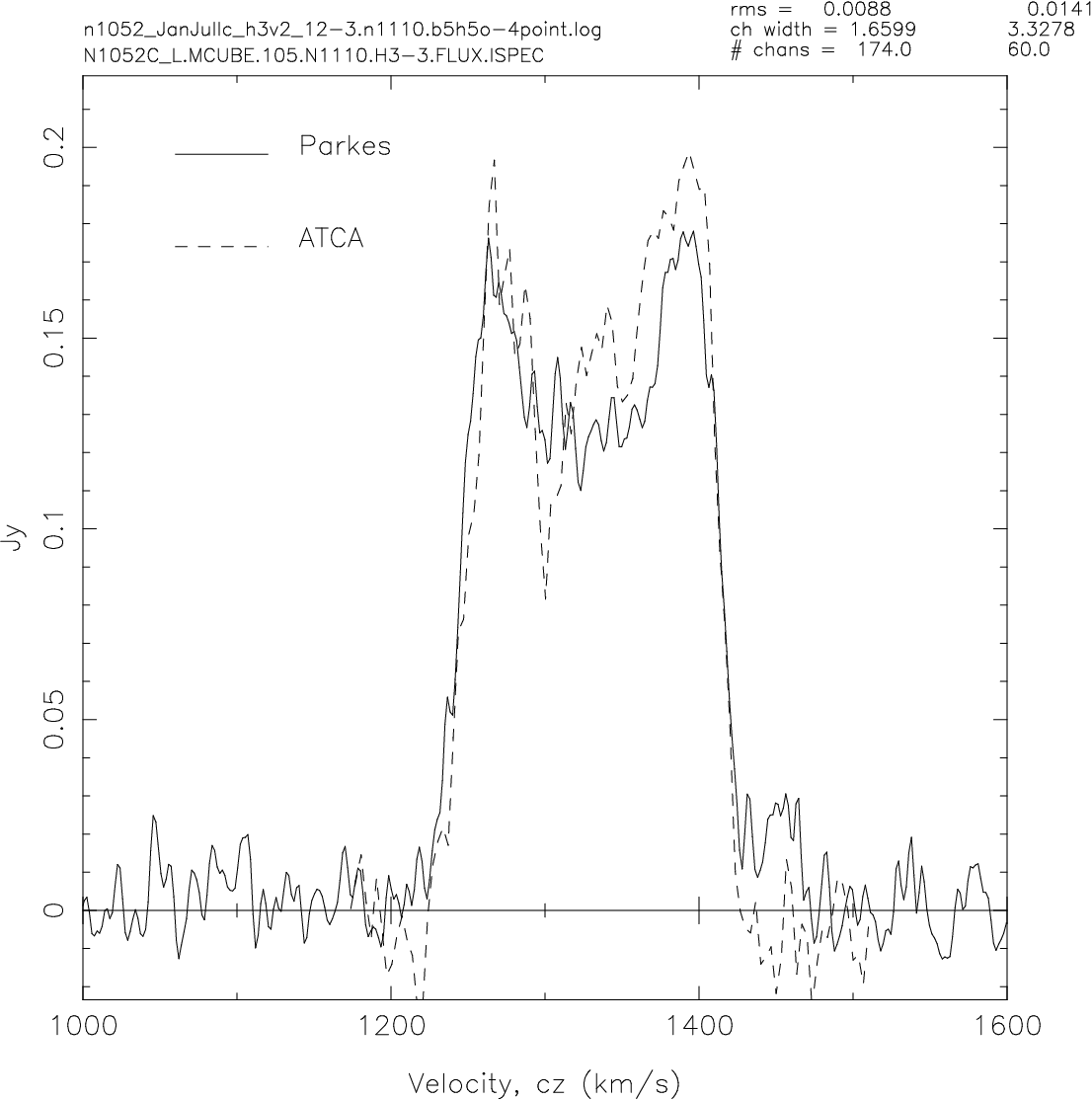}
}
\caption{NGC~1052 Group: Parkes (left) and ATCA (right) \hi{} spectra of
galaxy NGC~1110.
The Parkes spectrum was produced using a spatial box of width 5 pixels
(20 arcmin) and Hanning smoothing over 5 velocity channels. A
fourth-order polynomial has been used to correct the non-zero
baseline.  The vertical lines indicate the spectral region over which
the integral has been taken (see text, Section~\ref{sec:pks_1052}).
The ATCA spectrum was taken over a spatial region whose boundaries are
defined by an \hi{} column density of $1.57 \times 10^{20}$~atoms~cm$^{-2}$.}
\label{fig:n1110spectra}
\end{figure*}

Inspection of the Parkes cube for the NGC~1052 Group revealed \hi{}
emission from a number of known galaxies. In particular, our spectrum
of the edge-on galaxy NGC~1110 (ID~463 in \citealt{tc88}; RFGC~600 in
\citealt{kkk+99}), which lies $\sim$2\degrees{} from the group centre,
shows a classic line profile of a rotating disk
(Fig.~\ref{fig:n1110spectra}).  However, an additional emission
component distinct in velocity from NGC~1110 was detected, offset by
approximately one beamwidth south.  The spectrum taken at this
position, of the object hereafter named ``J$0249-0806$'', is shown in
Fig.~\ref{fig:n1052spectra}. A box width of 3~pixels and Hanning
smoothing over 5 channels were used to produce this spectrum, and a
fourth-order polynomial was fitted and subtracted to correct the
non-zero baseline.  The intensity integrated over velocity of
J$0249-0806$ is $\int\mathrm{S}~d\mathrm{V}=5.7\pm0.2$~Jy~km~s$^{-1}$,
implying an \hi{} mass of $\Mhi=(5.4\pm0.1)\times10^8\Msun$, assuming
a distance to the source of 20~Mpc.

The angular resolution of the Parkes data was not sufficient to
determine the spatial relationship between NGC~1110 and the
newly-discovered \hi{} emission. We therefore used ATCA to image the
region with an improved angular resolution 
($420\arcsec \x 50\arcsec$), which confirmed the new detection
and established this \hi{} emission as spatially and kinematically
distinct from that of NGC~1110.

\begin{figure*}
  \resizebox{!}{0.85\textheight}{
    \includegraphics*[16,71][442,770]{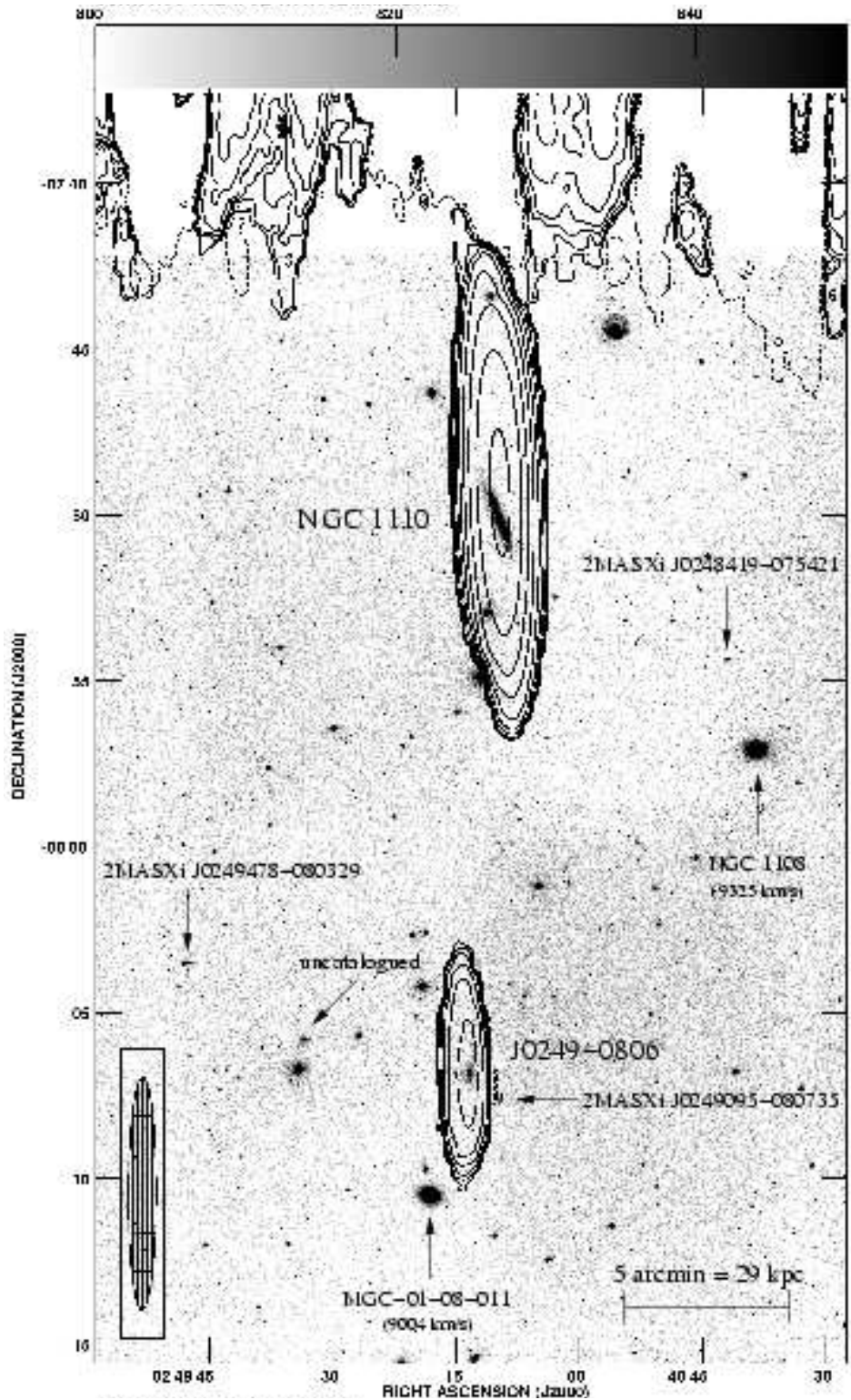}
  }
\caption{NGC~1052 Group: \hi{} total intensity contours of known galaxy
NGC~1110 (top) and new group member J$0249-0806$ (bottom), superimposed
on to the MPG/ESO 2.2~m WFI {\em R}-band optical image.
The peak column densities of the two sources are 
$5.4 \times 10^{21}$~atoms\,cm$^{-2}$ (NGC~1110) and
$1.1 \times 10^{21}$~atoms\,cm$^{-2}$ (J$0249-0806$).
The contour levels are 2, 4, 8, 16 and 32 times $1.3 \times
10^{20}$~atoms\,cm$^{-2}$ (2, 4, 8, 16 and 32~\Msun\ppcsq{}).
For presentation purposes the \hi{} map has been rescaled to the same
pixel size and coordinate grid as the optical data. The increase in the
background noise level at the top edges of the map are due to the
decrease in sensitivity of the primary beam with increasing distance
from the phase centre. 
Because the RMS noise is not uniform across the field, due to the
primary beam correction, noise measurements local to NGC1110 were used
for the moment analysis. 
The ellipse in the lower-left of the image indicates the size of the
synthesised beam ($420\arcsec \times 50\arcsec$).
The {\em R}-band image includes a number of background galaxies, 2MASS
galaxies with unknown velocities and uncatalogued galaxies. At the
calculated group distance of 20~Mpc, the projected linear scale of the
image is 1~arcsec~=~97~pc, represented by the scale bar at the
top-left of the figure.}
\label{fig:n1052_m0}
\end{figure*}

Fig.~\ref{fig:n1052_m0} shows the distribution of \hi{} emission from
the two galaxies, overlaid on our {\em R}-band optical
image. J$0249-0806$ lies 17~arcmin ($\sim95$~kpc)
south of the edge-on galaxy NGC~1110, and is associated with faint
diffuse optical emission. The \hi{} is centred on
\raJnf{02}{49}{14}, \decJnf{-08}{06}{51}, and has a velocity range 
1404--1490~\kms{}.  The rms noise level is $3.8$~\mJyb{} per
channel, with a $3\sigma$ detection corresponding to an \hi{} column
density of $\Nhi=3.0\times10^{18}$~cm$^{-2}$
(0.02~\Msun\ppcsq) per channel.
The \hi{} emission from NGC~1110 is spatially and kinematically
distinct from this, being centred on
\raJ{02}{49}{09}{7}, \decJnf{-07}{49}{17}, with velocity 
range 1237--1424~\kms{}.

\begin{figure*}
  \resizebox{\textwidth}{!}{
    \includegraphics*[17,71][550,634]{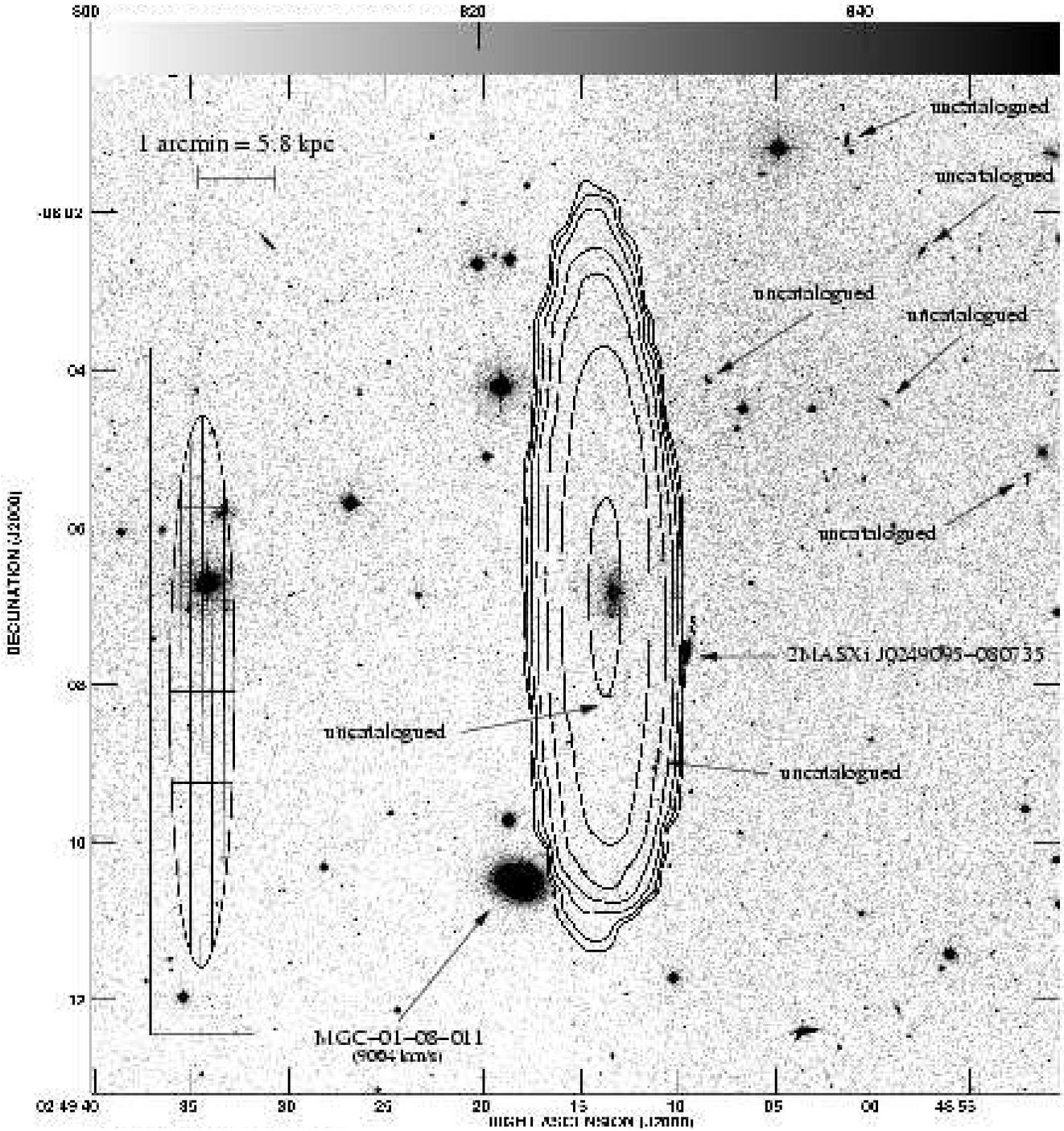}
  }
\caption{J$0249-0806$ in the NGC~1052 Group: Total \hi{} intensity
contours of the new group member superimposed onto the MPG/ESO 2.2~m
WFI {\em R}-band optical image. This is the same as
Fig.~\ref{fig:n1052_m0}, zoomed in on the area around J$0249-0806$,
and with background noise measured local to this source.
The peak column density of the source is 
$2.2 \times 10^{20}$~atoms\,cm$^{-2}$ (1.8~\Msun\ppcsq{}), with
contours representing \hi{} column density levels of 1, 2, 3, 4, 5, 6
and 7 times $1.3 \times 10^{20}$~atoms\,cm$^{-2}$ (1, 2, 3, 4, 5, 6
and 7~\Msun\ppcsq{}).
The \hi{} map has been rescaled to the same pixel size and coordinate
grid as the optical data.  The ellipse in the lower-left of the image
indicates the size of the synthesised beam 
($420\arcsec \times 50\arcsec$). 
At the calculated group distance of 20~Mpc, the scale of the image
is 1~arcsec~=~97~pc, indicated by the scale bar at the top-right of
the figure. Other objects in the field have been labelled, but
none of these is known to be associated with the group.}
\label{fig:n1052_j0249-0806_m0}
\end{figure*}

The total intensity map of J$0249-0806$ is shown in
Fig.~\ref{fig:n1052_j0249-0806_m0}.  This source is essentially
unresolved in \hi{} at the resolution of our observations, so little
may be determined about its morphology.  The (lower limit) peak \hi{}
flux is $2.2\times10^{20}$~atoms\,cm$^{-2}$ (equivalent to
1.8~\Msun\ppcsq{}) and lies close to the optical peak.
The \hi{} velocity structure shows a smooth change in velocity of
$\sim$40~\kms{} over $\sim 5$~arcmin, with marginal evidence of a
north-east to south-west velocity gradient with position angle
$\sim\degrees{30}$.  The \hi{} intensity of J$0249-0806$, integrated
over velocity, is
$\int\mathrm{S}~d\mathrm{V}=2.3\pm0.1$~Jy~km~s$^{-1}$.  This
corresponds to a total \hi{} mass
$\Mhi=(2.2\pm0.1)\times10^8$~\Msun{}, assuming the source to be at the
distance of the NGC~1052 Group.

The difference between the Parkes and ATCA velocity-integrated
\hi{} intensities for J0249-0806 (see Fig.~\ref{fig:n1052spectra}) implies
the presence of extended emission, not detected by the ATCA
observations due to the lack of very short antenna spacings, or the
presence of a significant amount of \hi{} with column densities below
the detection limit of the ATCA observations.
The map shown in Fig.~\ref{fig:n1052_m0} maximises the sensitivity to
extended structure (by only including data with
uv-distance~$<3.5\mathrm{k\lambda}$) available with the chosen array
configuration. This does not show any evidence of extended emission,
for example tidal tails, connecting bridge or common envelope
surrounding J$0249-0806$ and NGC~1110, which one might expect if the
two objects were strongly interacting. This lack of tidal features is
consistent with the current strength of the interaction, as defined by
the Dahari Q parameter (Dahari 1984), being very small.  We conclude
that the additional \hi{} emission detected in the Parkes spectrum of
J0249-0806 must be smoothly extended on scales larger than 4 arcmin
(the largest angular scale imageable by the array), possibly
distributed as a diffuse halo.

\subsubsection*{Optical data}

The MPG/ESO 2.2-m WFI {\em R}-band follow-up image of the region
clearly shows NGC~1110, and an uncatalogued low surface brightness
source, spatially coincident with the J$0249-0806$ neutral hydrogen
detection.  The image also shows a number of other galaxies (see
Figs~\ref{fig:n1052_m0} and~\ref{fig:n1052_j0249-0806_m0}) however
those with published velocities clearly lie outside the NGC~1052
group.  The derived optical properties of J$0249-0806$ are listed in
Table~\ref{tab:n1052_opt}.

\begin{table}
\caption{J$0249-0806$ in NGC~1052 Group: Measured and derived optical
properties. The estimate of Galactic extinction, A$_R$, has been
derived as per \citet{sfd98}. A$_R$(intrinsic) is the extinction
correction derived from the \hi{} content of J$0249-0806$.  Extinction
corrections have been applied to the absolute magnitude and
luminosity. Surface brightness, $\mu_R$, has been calculated for the
region within $D_{25}$.  }
\begin{tabular}{lc}
\hline
\multicolumn{2}{c}{J$0249-0806$} \\
\hline
m$_R$                           (mag) & $14.27$ \\ 
$\mathrm{A_R}$(Galactic)        (mag) & 0.084 \\
$\mathrm{A_R}$(intrinsic)       (mag) & 0.09 \\
$D_{25}$                        (arcsec) &  $13.6$ \\
M$_R$                           (mag) & -17.41 \\
$\mu_R$                         (mag arcsec$^{-2}$) & 22.1 \\
L$_R$                           (\Lsun) & $4.8 \times 10^8$ \\
\Mhi / L$_R$                    (\Msun/\Lsun) & 1.13 \\
\hline
\end{tabular}
\label{tab:n1052_opt}
\end{table}

\section{Discussion}
\label{sec:discussion}
\subsection{Group membership}
\subsubsection*{J1320--1427 and the NGC~5044 Group}

Early classification by \citet{hg82} of this galaxy group, using the
percolation method to determine group membership to a limiting
magnitude of m$_{\mathrm{B}} = 13.2$, suggested a membership of three
galaxies, concentrated around the brightest group member, NGC~5044.
\citet{gar93} subsequently used percolation and hierarchical
clustering \citep{mat78} methods to identify 9 group members to a
magnitude limit of m$_{\mathrm{B}} = 14.0$, and catalogued the group
as LGG~338. In a study of the central \degrees{2.3}sq. of the group,
which used the visual inspection of optical plates to find dwarf group
members, \citet{fs90} identified 80 definite and a further 82 likely
or possible members of the NGC~5044 Group, with 69\% of the total
classified as dwarf galaxies \citep{cel99}.

As part of the GEMS project, \citet{op04} have recalculated the
optical membership of the NGC~5044 Group, using NED as the parent
catalogue.  Their criteria for group membership require that members
lie within the $r_{500}$ overdensity radius of the group (calculated
from the group's X-ray temperature) and have a velocity which differs
by no more than $3\sigma$ from the group velocity.  Using this method
\citeauthor{op04} found 18 members of the NGC~5044 Group and
determined several group parameters, including the heliocentric
velocity $v = 2518\pm 100$~\kms{}, resulting in a corrected distance
to the group of 33~Mpc. The velocity dispersion,
$\sigma_{v} = 426\pm 74$~\kms{}, is high for galaxy groups.  These,
and other group parameters from \citet{op04} relevant to this paper
are given in Table~\ref{tab:group_params}.

\begin{figure*}
  \resizebox{\textwidth}{!}{
      \includegraphics[35,180][567,682]{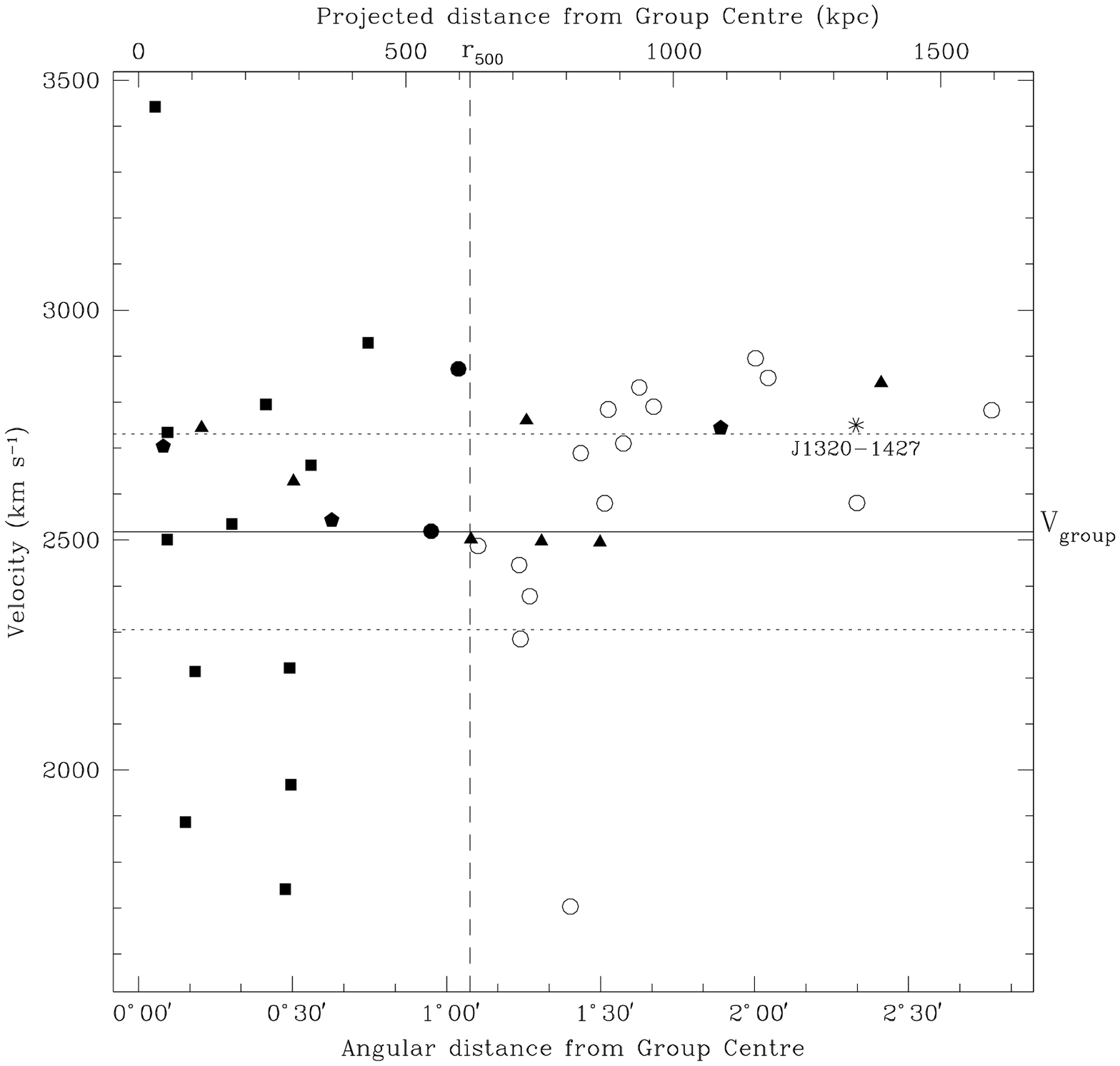}
  }
\caption{Velocity distribution of previously identified NGC~5044 Group 
members (solid symbols) and other probable group members within the
same projected area (open circles), as a function of angular distance
from the centre of the group. Symbols denote the galaxies in the group
in the same manner as in Fig.~\ref{fig:lgg338_pos}.  The new group
member reported in this paper, J$1320-1427$, is located in the right
of the diagram and is marked $\ast$. The group velocity is shown with
a solid horizontal line at $v = 2518$~\kms{}, while the dotted lines
parallel to this indicate the velocity dispersion of $\sigma =
426$~\kms. The dashed vertical line shows the extent of
$r_{500}$. Position and velocity data are from NED.}
\label{fig:lgg338_vel}
\end{figure*}

The position of J$1320-1427$ within the NGC~5044 Group is shown in
Fig.~\ref{fig:lgg338_pos}.  It lies on the outskirts of the group,
137~arcmin (\degrees{2.3}) north-west of the group centre.  This
corresponds to a projected distance of 1.3~Mpc, or $2.1r_{\mathrm{500}}$.
Despite its distance from the group centre, the velocity of J$1320-1427$
matches well with those of other group members. 
The velocity distribution of all known group members, as a function of
angular distance from the group centre, is shown in
Fig.~\ref{fig:lgg338_vel}; the velocity of J$1320-1427$ (2750~\kms{}) is
$1.1 \sigma$ from the mean group velocity of $2518 \pm 100$~\kms{}.  
We therefore suggest that J$1320-1427$ is likely to be a member of the
NGC~5044 Group, rather than a chance superposition of the group and an
unrelated field galaxy.

\subsubsection*{J0249--0806 and the NGC~1052 Group}
The NGC~1052 group of galaxies, also known as LGG~71, was first
identified by \citet{hg82} with a membership of 6~galaxies, with this
later extended to 14~galaxies by \citet{tul88} and \citet{gar93},
using the membership determination methods described above (see
Figs~\ref{fig:lgg71_pos} and~\ref{fig:lgg71_vel} for the position and
velocity distribution of the group members).
Using the same criteria employed for NGC~5044, \citet{op04} have
catalogued this group with 4~galaxy members, a group radial
velocity $v=1366\pm41$~\kms{} (implying a distance to the system of
20~Mpc) and velocity dispersion $\sigma_{v}=91\pm 35$~\kms{} (see
Table~\ref{tab:group_params}).

The brightest galaxy in the group is NGC~988 \citep{gmc+00}, an
SB(s)cd galaxy with m$_{\mathrm{B}} = 10.60$.  The galaxy which gives
the group its common name, NGC~1052, is classified as an E4 elliptical
\citep{rmg78} and is found to have an asymmetrical \hi{} spectrum.
\citet{vkr+86} suggest that the \hi{} gas was captured from a gas-rich
dwarf or spiral galaxy about $10^9$~years ago.

The radial distance of \degrees{2.1} between J$0249-0806$ and the
group centre corresponds to a projected distance of 747 kpc, or
$2.1r_{\mathrm{500}}$. The radial velocity of this galaxy,
1450~\kms{}, is $1.8\sigma$ from the mean group velocity of
$1366\pm41$~\kms{}.
Fig.~\ref{fig:lgg71_vel} shows the velocity distribution of members
of the NGC~1052 Group as a function of radial distance from the group
centre. From this, and the diagram showing the positions of the group
members on the sky (Fig.~\ref{fig:lgg71_pos}) we suggest that
J$0249-0806$ is a physical member of the NGC~1052 Group.

\subsection{The nature of J1320--1427 and J0249--0806}
The measured properties (\hi{} emission and optical photometry) of
J1320--\nolinebreak[4]1427 and J0249--0806 are consistent with those
exhibited by irregular galaxies.  

Morphologically, a smooth diffuse halo enveloping bright star forming
regions, akin to what we see in J$1320-1427$, is a feature occasionally
seen in high sensitivity observations of irregular galaxies
\citep{gh84}.
The absolute magnitude of this galaxy, measured here to be
M$_B=-16.54$, is in the range accepted for irregulars, tending
somewhat towards the dwarf end of that scale \citep{bst88,gh84}.

The mean \hi{} surface densities of the two galaxies, measured to a
radius where the \hi{} column density drops to 1~\Msun\ppcsq, are
$\sigmahi=2.11$~\Msun\ppcsq for J1320--1427 and
$\sigmahi=0.89$~\Msun\ppcsq for J0249--0806, with peak column
densities of 9.0\Msun\ppcsq and 1.8\Msun\ppcsq{} respectively. These
values are higher than the Kennicutt star-formation threshold
\citep{ken89}, and consistent with the surface density of late-type
spiral and irregular galaxies \citep{rh94}.

The neutral hydrogen mass-to-light ratios of J$1320-1427$ and
J$0249-0806$ are \Mhi/L$_R=2.56$, \Mhi/L$_B=1.64$ and \Mhi/L$_R=1.13$,
respectively while for the low surface brightness halo in J$1320-1427$
\Mhi/L$_R=1.05$~\Msun/\Lsun{} and \Mhi/L$_B=1.37$~\Msun/\Lsun{}.
The \Mhi/L$_B$ values are higher than the median values calculated by
\citet{rh94} (\Mhi/L$_B=0.36$~\Msun/\Lsun{} for Scd and Sd galaxies
and \Mhi/L$_B=0.66$~\Msun/\Lsun{} for Sm and Im galaxies) and others
\citep[eg.][etc.]{rob69, br97}, but they are consistent with several
individual Sm and Im galaxies in their samples, at the high end of the
\Mhi/L scale.
The \Mhi/L$_R$ values are consistent with the average for Im galaxies,
\Mhi/L$_R=1.92\pm0.71$, calculated from \hi{} masses and photometry by
\citet{sb02}.
In comparison, the isolated \hi{} cloud J1712-64, found by the HIPASS
survey \citep{ksm+00}, has \Mhi/L$_B=24$~\Msun/\Lsun{}, a further
indication, in addition to the fact that we do detect a stellar
component in both our objects, that J$1320-1427$ and J$0249-0806$ are
not primordial gas clouds left over from the formation of the groups.

Although not a primordial cloud, J$0249-0806$ might be on its first
infall into the NGC~1052 group.
As discussed in Section~\ref{sec:discussion}, this galaxy lies
approximately $2r_{500}$ from the group centre, but has a velocity
consistent with group membership. More indicative of first infall,
however, is its high \Mhi/L ratio, implying an absence of tidal
stripping. Aditionally, we have detected greater \hi{} emission with
the single-dish than in the synthesis observation, suggesting the
presence of gas extended on scales larger than 4~arcmin. This gas does
not appear to be in the form of tidal tails and therefore might exist
as a smooth diffuse halo, as is seen in, for example, Sextans A
\citep{wh02}.  Existence of such halos would confirm that the galaxy
is on its first infall into the group potential well, as otherwise
such a halo would easily have been stripped \citep[eg.][]{asp03,
gkk+02}.
In particular, simulations by \citet{mbd03}, of dwarf galaxies
moving through a low density IGM typical of galaxy groups, indicate
that the galaxies are completely stripped by the group IGM within
timescales as short as 1-2~Myr. An extended \hi{} halo, as may exist
around J$0249-0806$, is less likely to survive in the inner region of
the group, as it will be much less strongly bound than the disc gas
simulated by \citet{mbd03}.
Conversely, we see no evidence for first infall of the NGC~5044
group galaxy, J$1320-1427$.

\section{Conclusions}
We have presented high angular resolution radio synthesis and optical
observations of new galaxy members, discovered in the groups NGC~5044
and NGC~1052, as part of the GEMS survey of galaxy groups.

J1320--1427, in the NGC~5044 Group, has
$\Mhi=\nolinebreak1.05\x\nolinebreak10^9$~\Msun{} (single dish
measurement), $M_B=-16.54$, $M_R=-17.25$ and $\Mhi/L_B=1.64$~\Msun/\Lsun{}, 
$\Mhi/L_R=2.56$~\Msun/\Lsun{}.
J$0249-0806$, in the NGC~1052 Group, has $\Mhi=5.4\x10^8$~\Msun{}
(single dish measurement), $M_R=-17.41$ and $\Mhi/L_B=1.13$~\Msun/\Lsun{}.
Both of these galaxies are at a distance equivalent to just over
$2r_{500}$ from their group centres. This and the possible presence of
an extended \hi{} halo suggests that J$0249-0806$ may be on its first
infall into the NGC~1052 group.

\section*{Acknowledgments} 
Financial support in the form of postgraduate studentships from the
Astrophysics Research Institute and PPARC is acknowledged by NPFM and
SB respectively.
NPFM would like to thank the Institute of Astrophysics at the
University of Innsbruck and ESO Garching for their hospitality on
several occasions whilst this work was being carried out.
CGM acknowledges financial support from the Royal
Society. 
The authors would also like to thank the referee, for useful comments
regarding this work.
This research has made use of the NASA Astrophysics Data
System Abstract Service (ADS), and the NASA/IPAC Extragalactic
Database (NED), which is operated by the Jet Propulsion Laboratory,
California Institute of Technology, under contract with the National
Aeronautics and Space Administration.



\begin{figure*}
  \resizebox{\textwidth}{!}{
    \rotatebox{-90}
    {
      \includegraphics{mckay-figure02.ps}
    }
  }
\caption{Member positions in the NGC~1052 Group. 
Solid symbols show the position of previously identified group
members, whilst open circles denote the positions of other probable
members (catalogued by NED) within the same projected area and with
velocities close to that of the group.
The symbol representing each member denotes the first group
catalogue to list the galaxy:
pentagon--\citet{hg82}, 
square--\citet{tul88}, 
triangle--\citet{gar93}.
The position coordinates are as given by NED.
The centre of the group lies close to the galaxy NGC~1052 itself, at
\raJ{02}{40}{35}{3}, \decJnf{-08}{13}{08}. The extent of the 
$r_{500}$ overdensity radius is shown with a dashed circle centred on
this point.  The new group member reported in this paper,
J$0249-0806$, is in the centre left of the diagram (marked~$\ast$)
and is surrounded by a circle indicating the size of the ATCA field of
view. The position of the earby galaxy, NGC~1110 is also labelled.
}
\label{fig:lgg71_pos}
\end{figure*}

\begin{figure*}
\resizebox{\textwidth}{!}{
  \includegraphics*[40,24][534,527]{mckay-figure09.ps}
}
\caption{NGC~1052 Group: Parkes and ATCA \hi{} spectra of new group member
J$0249-0806$.
The Parkes spectrum was produced using a spatial box of width 3 pixels
(12 arcmin) and Hanning smoothing over 5 velocity channels. A
fourth-order polynomial has been used to correct the non-zero
baseline.  The vertical lines indicate the spectral region over which
the integral has been taken (see text, Section~\ref{sec:pks_1052}).
The ATCA spectrum was taken over a spatial region whose boundaries are
defined by an \hi{} column density of $1.57 \times 10^{20}$~atoms~cm$^{-2}$.
}
\label{fig:n1052spectra}
\end{figure*}

\begin{figure*}
  \resizebox{\textwidth}{!}{
    \includegraphics[35,180][567,682]{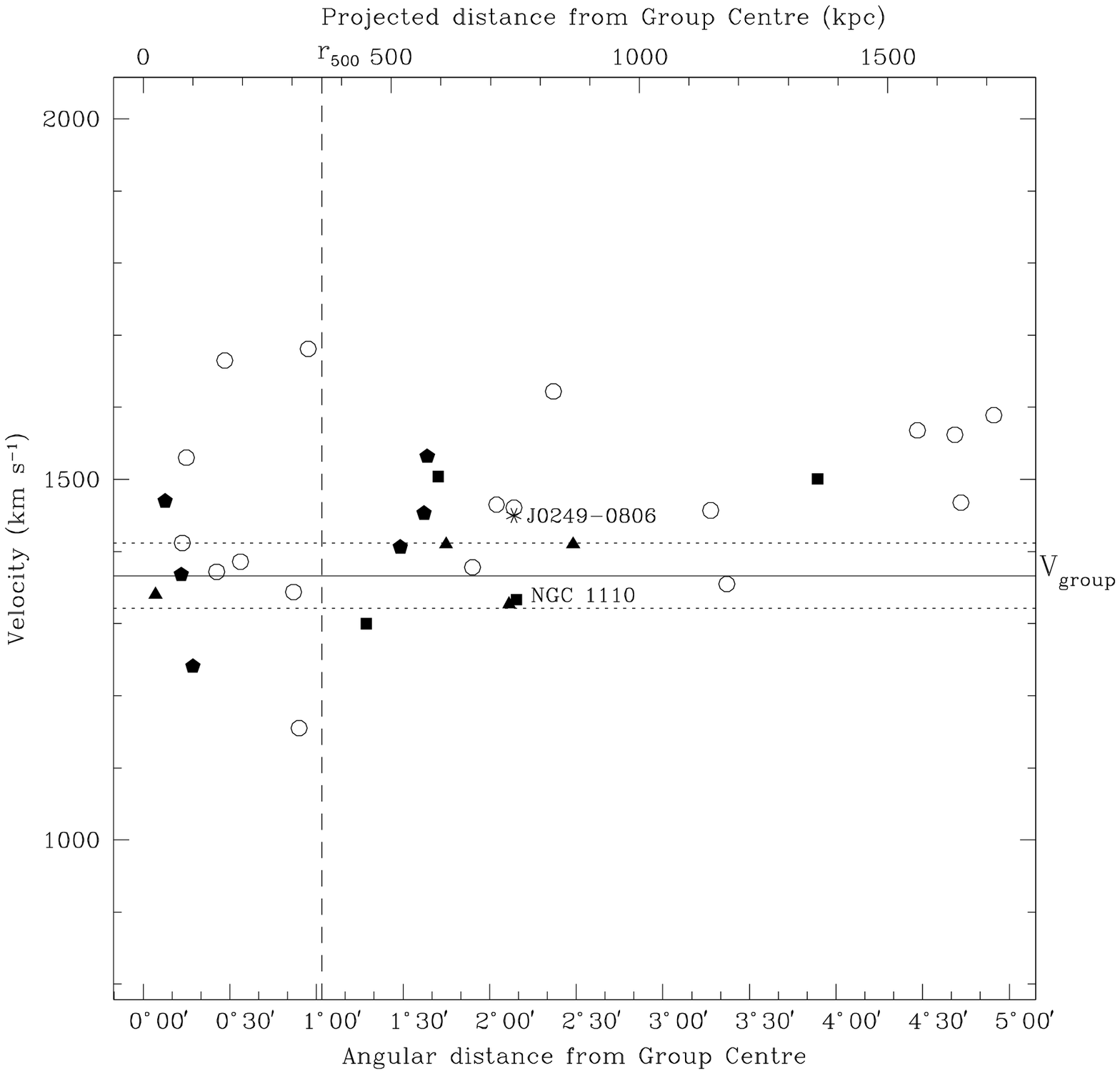}
  }
\caption{Velocity distribution of previously identified NGC~1052 Group 
members (solid symbols) and other probable group members within the
same projected area (open circles), as a function of angular distance
from the centre of the group.  Symbols denote the galaxies in the
group in the same manner as in Fig.~\ref{fig:lgg71_pos}.  The new
group member reported in this paper, J$0249-0806$, is located just
above centre in the diagram and is labelled and marked $\ast$. The
nearby galaxy NGC~1110 is also labelled and marked with a filled
square, for comparison.  The group velocity is shown with a solid
horizontal line at $v = 1366$~\kms{}, while the dotted lines parallel
to this indicate the velocity dispersion of $\sigma = 91$~\kms.  The
dashed vertical line represents the extent of $r_{500}$.  Position and
velocity data are from NED.  }
\label{fig:lgg71_vel}
\end{figure*}

\clearpage
\clearpage

\end{document}